\documentclass[12pt,epsf]{article}
\usepackage{graphicx, amsmath}

\begin{document}

\sloppy
\begin{flushright}{SIT-HEP/TM-26}
\end{flushright}
\vskip 1.5 truecm
\centerline{\large{\bf Primordial black holes from monopoles connected by strings}}
\vskip .75 truecm
\centerline{\bf Tomohiro Matsuda
\footnote{matsuda@sit.ac.jp}}
\vskip .4 truecm
\centerline {\it Laboratory of Physics, Saitama Institute of
 Technology,}
\centerline {\it Fusaiji, Okabe-machi, Saitama 369-0293, 
Japan}
\vskip 1. truecm
\makeatletter
\@addtoreset{equation}{section}
\def\theequation{\thesection.\arabic{equation}}
\makeatother
\vskip 1. truecm

\begin{abstract}
\hspace*{\parindent}
Primordial black holes (PBHs) are known to be produced from collapsing
 cosmic defects such as domain walls and strings. In this paper we show
 how PBHs are produced in monopole-string networks.
\end{abstract}

\newpage
\section{Introduction}
\begin{figure}[ht]
 \begin{center}
\begin{picture}(400,380)(0,0)
\resizebox{11cm}{!}{\includegraphics{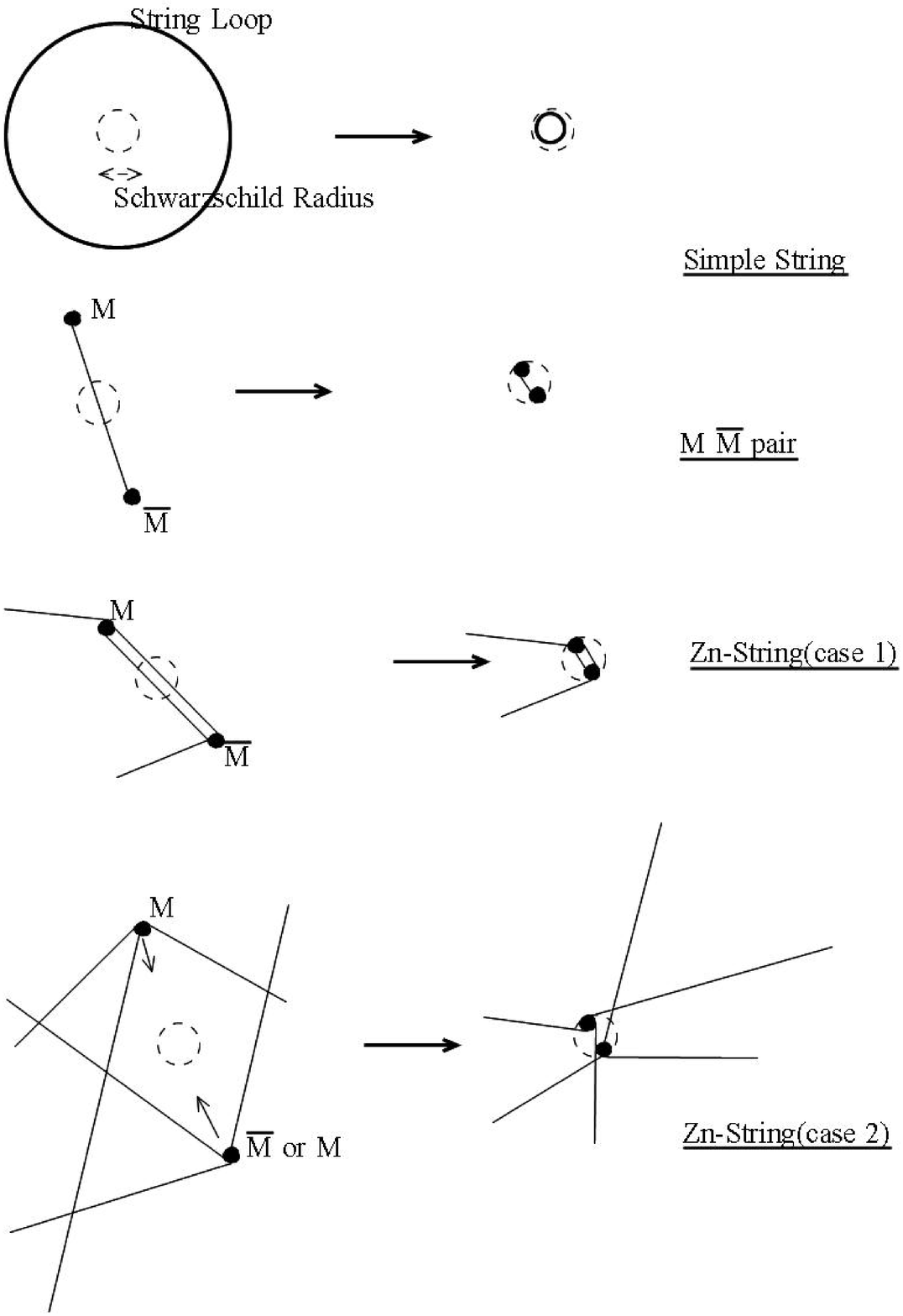}} 
\end{picture}
\caption{The figure in the first line shows an example of 
the PBH formation in Hawking's scenario. 
We show the simplest example of a perfect circle. 
The figure in the second line shows the PBH formation from a pair of
monopoles.
The figures in the third and the last line shows examples of the
  network of $Z_3$ strings.
Our model of the PBH formation is a natural extension of the original
scenario in the sense that the kinetic energy of the defect at the
  collision plays an important role.}
\label{fig1}
 \end{center}
\end{figure}

S.W.Hawking has discussed how cosmic string loops that shrink by a
factor of order $1/G\mu$ will form black
holes\cite{Hawking}\footnote{Other possible origins and applications are
discussed in ref.\cite{other_PBH}.}. 
In this case, a tiny fraction $f\ll 1$ of order
$(G\mu)^{2x-4}$, where $x$ is the ratio of the loop length to the
correlation length, loops will form black holes\footnote{See the first
picture in Fig.\ref{fig1}.}. 
The result obtained by Hawking is cosmologically important because the
emission of $\gamma$-rays from little black holes is
significant\cite{post-Hawking}.
Numerical simulation of loop fragmentation and evolution was studied
later by Caldwell and Casper\cite{PBH_numerical}, where the authors
obtained the value of the fraction; 
\begin{equation}
\label{fraction}
f=10^{4.9\pm0.2}(G\mu)^{4.1\pm0.1}.
\end{equation}
Black holes created by these collisions are so small that they lose
their energy due to the Hawking evaporation process. 
The fraction of
PBHs today in the critical density of the Universe is discussed by
MacGibbon et al\cite{branden}, where the authors calculated
the fraction of black hole remnants 
\begin{equation}
\Omega_{PBH}(t_0) = \frac{f}{\rho_{crit}(t_0)}\int^{t_0}_{t_*}
dt \frac{dn_{BH}}{dt}m(t,t_0),
\end{equation}
where $t_0$ is the present age of the Universe, and $t_*$ is the time
when PBHs with initial mass $M_* \simeq 4.4\times 10^{14} g$ were formed
and which are expiring today.
$m(t,t_0)$ is the present mass of PBHs created at time $t$.
The approximate form of the mass function $m(t,t_0)$ is given by
\begin{equation}
\label{mat-hawk}
m(t,t_0)\simeq \alpha \mu t.
\end{equation}
The extragalactic $\gamma$-ray flux observed
at 100MeV is commonly accepted as providing a strong constraint on the
population of black holes today.
According to Carr and MacGibbon\cite{Gibbons-carr}, the limit implied by
the EGRET experiment is
\begin{equation}
\Omega_{PBH} <  10^{-9}.
\end{equation}
The scaling solution of the conventional string network suggests that the
rate of the formation of PBHs is
\begin{equation}
\label{string_BH}
\frac{d n_{BH}}{dt} = f\frac{n_{loop}}{dt}\sim \alpha^{-1}f t^{-4}.
\end{equation}
Using the above results one can obtain an upper bound\cite{branden}
\begin{equation}
G\mu < 10^{-6},
\end{equation}
which is close to the constraint obtained from the normalization of the
cosmic string model to the CMB.

In addition to the simplest mechanism that we have stated above, it is
always important to find a new mechanism for PBH formation, especially
when PBHs produced by the new mechanism have a distinguishable property.
Based on the above arguments we will consider
less simplified networks of hybrid defects.
The networks we consider in this paper are;
\begin{enumerate} 
\item Pairs of monopole-antimonopole that are connected by strings.
\item Tangled networks of monopoles and strings where $n>2$ strings
      are attached to each monopole.
\end{enumerate}
We will show how PBHs are formed in the monopole-string networks.
There are qualitative and quantitative differences between our new
mechanisms and the conventional ones.

Before discussing the collision of the monopoles, it is
important to note that we are considering heavy PBHs that can 
survive until late and may affect our present Universe. 
This means that the separation must be large enough 
(i.e. $t_{in}$ must be late enough) so that the pairs can form ``heavy''
PBHs. 
In this case, unlike the conventional scenario of the
monopole-antimonopole ``annihilation'', 
the strings connecting the pairs do not have to dissipate their energy
before the gravitational collapse, since the monopoles that go into the
Schwarzschild radius cannot come back.
One might think that if there are monopoles that have unconfined charges 
other than the magnetic charge, which might happen in natural setups,
there could be a monopole-antimonopole 
``scattering'' occurring before they go into the Schwarzschild radius
and prevents the gravitational collapse.To get a rough understanding of the interactions mediated by a massless
gauge boson, it is helpful to remember a famous result
that the cross section for a scattering of the relativistic particles with
significant momentum is given by\cite{vilenkin_book}
\begin{equation}
\sigma \sim \frac{e^2}{T^2},
\end{equation}
where $T$ denotes the typical particle momentum.
From this equation people might think that the scattering
must prevent the gravitational collapse.
However, the monopole-antimonopole scattering cross section should not be
estimated by using the typical momentum of the particles in the thermal
plasma, but by the specific momentum of the colliding objects. 
In our model, the typical momentum of the monopoles
that are about to collide is obviously very large.
Scattering is not important in this case, since the cross section is
 much smaller than the Schwarzschild radius.

\section{Monopole-antimonopole connected by a string}
\begin{figure}[ht]
 \begin{center}
\begin{picture}(400,380)(0,0)
\resizebox{11cm}{!}{\includegraphics{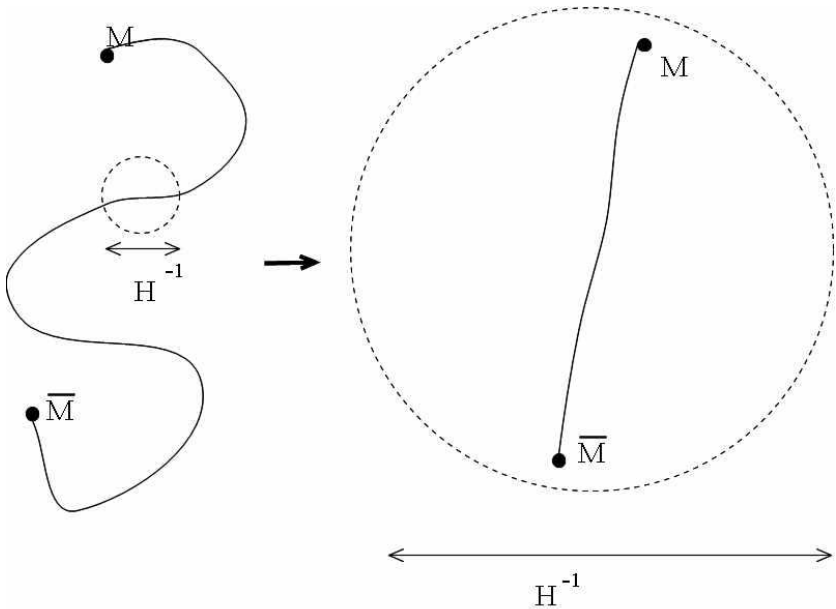}} 
\end{picture}
\caption{This picture shows an interesting possibility that may 
arise in the context of inflationary scenario\cite{vilenkin_book}.
Monopoles are formed during inflation but are not
completely inflated away.
Strings are formed by the succeeding phase transition that induces
confinement. 
Strings can either be formed later during inflation or in the
  post-inflationary epoch.  
A string that connects a pair is initially much longer than the Hubble
  radius.
Therefore, the strings connecting monopoles have Brownian shapes, as is
  shown on the left.
During the evolution, the correlation length of the strings
 grows faster than the monopole separation due to the small loop
  production and the damping force acting on the strings. 
Finally, the correlation length of the strings becomes
comparable to the monopole separation, and thus, one is left with
a pair of monopoles connected by more or less straight strings, as is
  shown on the right.}
\label{fig2}
 \end{center}
\end{figure}

First, we will consider a simple toy model.
Here we consider a model proposed by Langacker and
Pi\cite{Langacker-Pi}, in which the
Universe goes through a phase transition with the $U(1)$ symmetry of
electromagnetism spontaneously broken.
The most obvious consequence of this additional phase transition is
that during this phase monopoles and 
antimonopoles are connected by strings, due to the superconductivity
of the vacuum.
Let us first examine whether PBHs are formed in the original
Langacker and Pi scenario.
The separation between monopoles at the time of the 
string formation ($d(t_s)$) is bounded by\cite{vilenkin_book} 
\begin{equation}
\label{largest_d}
d(t_s) < (t_s t_M)^{1/2},
\end{equation}
where $t_M$ is the time when monopoles are produced.\footnote{Besides
the pairs that follow the conventional bound 
(\ref{largest_d}) in Ref.\cite{vilenkin_book}, there will be a network
of longer strings that has the scaling solution.
This network will remain after the conventional annihilation and will
lead to another kind of PBH formation. See also Section 3.}    
Strings are formed later at $t_s > t_M$, when charges are confined and 
monopoles are connected by the strings.
The total energy of a pair is about $\sim \mu d(t_s)$.
The Schwarzschild radius for this mass is given by
\begin{equation}
R_g \sim G \mu d.
\end{equation}
Here we can ignore the frictional forces acting on this system, since 
they do not alter the above result\cite{vilenkin_book}.
In this case, black holes are formed if the Schwarzschild radius is
larger than the width of the strings $\sim \eta^{-1}\sim \mu^{-1/2}$.
As was discussed by Hawking in ref.\cite{Hawking}, topological defects
that can shrink instantly to their Schwarzschild radius $R_g$
turn into black holes.
In our present model, the size of the monopoles is smaller than the
width of the strings, and they are connected by almost straight
strings\footnote{See \cite{vilenkin_book} for conventional reviews of
cosmic strings and monopoles. Keep in mind that we are considering
a very common situation.}.
In this case, the minimum size of the defect when they 
shrink is larger than the size of the monopoles but is 
about the same order as the width of the strings.
The condition for the black hole formation is therefore given by
\begin{equation}
\label{H-crit1}
R_g > \eta^{-1}.
\end{equation}
Unfortunately, the bound (\ref{largest_d}) contradicts 
the criteria given in eq.(\ref{H-crit1}). 
In order to increase the typical mass of monopole-string-antimonopole,
we will consider a less simplified model in which monopoles are diluted
but not completely inflated away during the 
period of an (additional) inflationary expansion.
The basic idea of the model is shown schematically in
Fig.\ref{fig2}.
As we have discussed in the previous section, black holes will be
formed when the monopole-antimonopole pair comes into the horizon
at $t=t_{in}$, where $t_{in}$ is determined by the dilution mechanism.
The production probability of PBHs is expected to be $O(1)$ for long
and straight strings that has monopoles at their endpoints.

The number distribution of the mass of the PBHs has a sharp peak.
One may find a similar characteristic in the conventional mechanism of
the PBH formation in models of hybrid inflation\cite{BLW}. 
The typical mass of such PBHs is given by\cite{BLW}
\begin{equation}
\label{pbh_hybrid}
M_{pbh}\simeq \frac{M_p^2}{H_I}e^{2N_c},
\end{equation}
where the Universe is assumed to be inflated $e^{N_c}$ times
after the phase transition.
$M_{pbh}$ in eq.(\ref{pbh_hybrid}) is about the same order as the total
mass that is contained in the horizon. 
To show explicitly the difference between the new model and the
conventional one, let us evaluate the typical mass (and the size) of 
PBHs produced in the monopole-string networks.
At the end of inflation, the physical distance between the diluted
monopoles is about 
$H_{I}^{-1}e^{N_c}$, where $H_I$ is the Hubble constant during
inflation.
Then, the scale factor of the Universe develops as 
$(t H_I)^{1/2}$, which is nothing but the usual evolution of the
radiation-dominated Universe.\footnote{We assume for simplicity that the
equation of state becomes $p=\rho/3$ (ultrarelativistic gas) soon after inflation.}
Then the typical distance between the diluted monopoles is given by
\begin{equation}
H_{I}^{-1}e^{N_c} \times (t H_I)^{1/2},
\end{equation}
which becomes comparable to the particle horizon at $t_{in} \sim
H_I^{-1} e^{2N_c}$. \footnote{We are assuming that the initial
separation of the monopole and antimonopole is larger than the Hubble
radius, which is due to the inflationary expansion after the phase
transition. We also assume that the distance will expand at constant
comoving distance until it enters the Hubble radius.} 
At this time, the mass of a monopole-antimonopole pair connected by a
string is 
\begin{equation}
\label{pbh_1st}
M_{pbh}\simeq \frac{\mu}{H_I}e^{2N_c},
\end{equation}
where $\mu$ is the tension of the string.\footnote{In spite of the
similarity between (\ref{pbh_1st}) and  (\ref{pbh_hybrid}), the origin
of the factor $e^{2N_c}$ is rather different. See ref.\cite{BLW} for more
details.}

\section{Tangled networks of monopoles and strings }
In the previous section we showed that PBHs are produced
in the monopole-string networks if inflation dilutes the monopoles. 
Therefore, it will be very interesting to consider a more complicated
model of 
the tangled monopole-string networks, where monopoles are connected to
$n>2$ strings.
According to the previous work in this field, the network is
characterized by a single length scale, 
$d(t) \sim \gamma t$\cite{vilenkin_book}.

Let us consider the networks of $Z_n$-strings\cite{bere-mar-vilenkin,
Vachaspati_Vilenkin,  Zn_other}. 
The first stage of symmetry breaking occurs at a scale $\eta_m$,
when monopoles are produced.
Then the second symmetry breaking produces a string network at a
scale $\eta$, where the symmetry breaking is given by
\begin{equation}
G\rightarrow K \times U(1) \rightarrow K\times Z_n.
\end{equation}
The monopole mass and the string tension are given approximately by
$m\sim 4\pi \eta_m/e$ and $\mu \sim \eta^2$ with gauge coupling $e$.
The evolution of the string-monopole network has been studied by
Vachaspati and Vilenkin\cite{Vachaspati_Vilenkin}.
These authors showed that the networks exhibit scaling behavior
\begin{equation}
\label{monopole-string}
d(t) \sim \gamma t,
\end{equation}
where $\gamma$ was taken from Berezinsky et al\cite{bere-mar-vilenkin}.
Assuming that radiation of gauge quanta is the dominant energy loss
mechanism of the monopole-string networks, the value of
$\gamma$ is calculated\cite{vilenkin_book, bere-mar-vilenkin}, and
is given by
\begin{equation}
\gamma \sim 4\pi\mu/e^2 m^2.
\end{equation}
In this model, the energy loss mechanism is important in determining 
this parameter.
In ref.\cite{bere-mar-vilenkin},  the authors discussed how the
ultrarelativistic motion of the monopoles may produce ultra-high energy
gamma rays.
Let us explain why monopoles in the networks of $Z_n$-strings may reach 
huge kinetic energy.
In monopole-string networks, with $n$ strings attached to each
monopole, the proper acceleration of a monopole should be determined by
the vector sum of the tension forces exerted by the $n$ strings, which
is given by $a\sim \mu/m$ by order of magnitude.
Therefore, considering the result (\ref{monopole-string}), one can 
 obtain the typical energy of a monopole\cite{bere-mar-vilenkin}
\begin{equation}
E_m \sim \mu d\sim \mu \gamma t.
\end{equation}
Now it is clear that the monopoles have huge kinetic
energy proportional to the separation distance and this distance 
grows with time.
What we consider in this paper is the gravitational collapse of such
monopoles.
Of course, the number of the collisions of monopoles per unit time 
is very small because of their small number density.
However, once PBHs are formed, they can be cosmologically important
even if their number density is very small.
Therefore, in the typical collision of such monopoles, one must not
disregard the black hole formation.
In this case, the typical mass of the black hole is given by
\begin{equation}
m_{BH}\sim \mu \gamma t.
\end{equation}
One may think that the above result looks similar to result
(\ref{mat-hawk}) obtained for the simplest string networks.
However, remember that the networks that we are considering in
this section are quite different from the conventional string networks.
We are not considering the PBH formation from string loops, but 
the ones that come about from the collision of the energetic monopoles
that come closer, within their Schwarzschild radius, $R_g \sim G\mu
\gamma t$. 

Let us calculate the number density of PBHs and see if we can put
bounds on the tension of the strings (or on the mass of the monopoles).
The number density of the monopoles is given by
$n_m \sim d^{-3}$.
Due to the random motion of the monopoles, the nucleation rate of the
black holes is given by the conventional formula which is
\begin{equation}
\label{numbe}
\frac{d n_{BH}}{dt}\sim n_m^2 \pi R_g^2 
\sim \frac{\pi \mu^2}{M_p^4 \gamma^4 t^4},
\end{equation}
where the velocity of the energetic monopoles is $v_m \simeq 1$.
Now it is easy to calculate the number distribution of the PBHs,
$dn_{PBH}/dM$.
Neglecting the mass loss of the black holes with the initial mass greater
than $M_*$, we obtained the present value of $dn_{PBH}/dM$ by
redshifting the distribution\cite{branden}. 
Following the conventional calculation\cite{branden}, 
the result is given in a straightforward manner  by
\begin{equation}
\frac{d n_{BH}}{dM} \propto M^{-2.5}.
\end{equation}
The number distribution of the PBHs obtained above looks similar to the
one 
obtained in Hawking's scenario.
However, there is a crucial discrepancy in the formation probability
of the PBHs, which is usually denoted by ``$f$'' in Hawking's
scenario. 
In the conventional scenario, the value of $f$ is given by
eq.(\ref{fraction}), which was obtained in ref.\cite{post-Hawking}.
The value of $f$ is $f\sim 10^{-20}$ for $G\mu\sim 10^{-6}$, which is of
course quite tiny and characterizes Hawking's mechanism.
In our case, however, ``$f$'' must be different from the ``$f$'' in 
Hawking's scenario because the situation of the PBH formation is
qualitatively different.
PBHs are formed whenever the energetic monopoles come closer than their
Schwarzschild radius.
The Schwarzschild radius is much larger than the size of the
monopoles in that we are considering large PBHs with mass
$M_{pbh}>M_*$. 
Therefore, the production probability of the colliding
monopoles is $f\sim 1$ for the heavy PBHs that can survive today, and
thus, $f$ in our scenario is much larger than the one obtained in
Hawking's scenario. 
On the other hand, the nucleation rate of PBHs is not about 
$10^{20}$ times as large as the conventional value, since there is a
small factor in eq.(\ref{numbe}), as we have discussed above.
Eq.(\ref{numbe}) may correspond to the
production ratio of the closed string loops in the conventional scenario.

Comparing our result with (\ref{string_BH}) and using the result
obtained by MacGibbon et al\cite{branden}, we obtained the constraint
\begin{equation}
(G\mu) < 10^{-10} 
\left[\frac{M_*}{4.4\times 10^{14}g} \right]^{1/7}
\left[\frac{\gamma}{10^{-2}}\right]^{5/7} 
\left[\frac{t_{eq}}{3.2\times 10^{10}s}\right]^{-1/7},
 \end{equation}
where the calculation is straightforward.
The result obtained here puts a new bound on the tension
of the strings.
However, the constraint is not important here.
What is important in this paper is that we have found
a novel mechanism for PBH formation, which is both qualitatively and
quantitatively different from the old ones.

\section{Conclusions and discussions}
For usual cosmic strings, we know that
only a tiny fraction of string loops can collapse to form black holes.
Although the fraction of the production probability is very tiny for
the conventional strings, this
unique mechanism for PBH formation is cosmologically very important.
In this paper, we considered two scenarios for the PBH formation
in the monopole-string networks.
The typical mass and the number distributions of the PBHs obtained in our
model are distinctive.

First, we considered a model 
in which monopoles are diluted(weakly inflated away) but not completely
inflated away. 
We found a narrow mass range and obtained $M_{pbh} \simeq
\frac{\mu}{H_I}e^{2N_c}$. 
Qualitatively, PBHs formed in our model looks similar to the ones
produced during hybrid inflation.
However, there is a hierarchical discrepancy in the typical mass.
The difference is due to the crucial differences between
the two mechanisms.

Our second model is the monopole-string networks with $n>2$ strings 
attached to each monopole.
We examined another mechanism of the PBH formation and 
found that the number density distribution 
 of the PBHs of mass $M$ is proportional to $M^{-2.5}$.
It is known that there is a similar distribution in the
 conventional string networks.
However, in the Hawking's scenario there is always a tiny probability $f$,
 which is absent in our model.
On the other hand, there is another small factor in our result.
The difference is due to the qualitative differences between the two
distinctive models, as we have discussed.
It may be important to note that our mechanisms can work in a hidden
sector.

In a previous paper\cite{matsuda_futu} we have considered cosmic
necklaces and discussed another important implications of the
defect-induced PBH formations, focusing our attention to brane
inflation in the brane Universe.
Although it has long been believed that ``only strings are produced in
the brane Universe'', it is not difficult to show explicitly how 
defects other than strings are produced in the brane Universe.
For example, monopoles, necklaces and domain walls are discussed in
ref.\cite{matsuda_mono_domain}, and Q-balls are discussed in
ref\cite{matsuda_Q-balls}.
A natural solution to the domain wall problem in a typical supergravity
model is discussed in ref.\cite{matsuda_wall}, where the required
magnitude of the gap in the quasi-degenerated vacua is induced by $W_0$
in the superpotential.
The mechanism discussed in ref.\cite{matsuda_wall} is natural since the
constant term $W_0$ in the superpotential is necessary so as to cancel
 the cosmological constant.
Cosmic strings and other defects are important because they are produced
after various kinds of brane-motivated inflationary models.
They could be used to
distinguish the brane models from the conventional
ones\cite{matsuda_futu}. 
Moreover, if the fundamental scale of the brane world is very low,
one needs to construct mechanisms of inflation and baryogenesis that
may work in the low-scale models.
The ideas of the low-scale inflationary models and baryogenesis are
discussed in ref.\cite{low-infla} and ref.\cite{low-baryo}, where
defects play crucial roles. 

\section{Acknowledgment}
We wish to thank K.Shima for encouragement, and our colleagues at
Tokyo University for their kind hospitality.

\end{document}